\documentclass[twocolumn]{revtex4}
\usepackage{epsfig}
\usepackage{graphicx}
\usepackage[tbtags]{amsmath}

\begin{document}

\title{Direct measurement of concurrence via visibility in a cavity QED system}
\author{Sang Min Lee$^{1}$, Se-Wan Ji$^{1}$, Hai-Woong Lee$^{1}$ and M. Suhail Zubairy$^{2,3}$}
\address{$^{1}$Department of Physics, Korea Advanced Institute of Science and Technology, Daejeon 305-701, Korea\\
$^{2}$Department of Physics and Institute for Quantum Studies, Texas A\&M University, College Station, Texas 77843-4242, USA\\
$^{3}$Texas A\&M University at Qatar, Education City, P.O.Box 23874, Doha, Qatar}

\begin{abstract}
An experimental scheme is proposed that allows direct measurement of the concurrence of a two-qubit cavity system. It is based on the cavity-QED technology using atoms as flying qubits and relies on the identity of the two-particle visibility of the atomic probability with the concurrence of the cavity system. The scheme works for any arbitrary pure initial state of the two-qubit cavity system.
\end{abstract}

\maketitle

The question of how to detect the presence and amount of
entanglement is one of the central issues in quantum information science. There
exist theoretical criteria and measures such as the positive partial
transpose (PPT) criterion \cite{1,2}, entanglement of formation
\cite{3} and concurrence \cite{4} that, in principle, allow one to
determine the presence and amount of entanglement. It is, however,
difficult to observe such criteria and measures experimentally. The PPT criterion involves a nonphysical operation of partial transposition (complex conjugation) of the density matrix elements, while the entanglement of formation and the concurrence are complicated nonlinear functions of the system state. One is thus led to think that one may have to rely on
the technique of a full tomographic reconstruction of the quantum
state to measure the entanglement of an unknown quantum state. This
technique, although successfully implemented for small systems
\cite{5,6}, is highly inefficient and difficult especially for large
systems, as a large number of observables need to be measured. The
question naturally arises whether entanglement can be estimated
without having to fully reconstruct the unknown state. It has been
shown that the answer to this question is yes at least for the case
of pure two-qubit states \cite{7,8}, although, even in this case,
more than one observable need to be measured.

In recent years, several methods \cite{9,10,11,12,13} have been
proposed for detecting and measuring entanglement without a full
reconstruction of the state; e.g., the method \cite{9} based on the technique of minimal and optimal tomography \cite{14,15} performed on one of the entangled pair, the method \cite{10} based on
entanglement witness \cite{16,17} which was realized experimentally
\cite{18}, the method \cite{11,12} based on PPT criterion
\cite{1,2}, and the method \cite{13} based on two-particle
interferometry \cite{19, 20}. These methods, although much simpler
than the full state reconstruction, are not completely free of
experimental difficulties, as they require either controlled unitary
operations or some prior knowledge about the quantum state in question,
or they can detect entanglement but not measure its amount.

 Very recently, direct measurement  of the concurrence of a two-photon pure entangled state was demonstrated experimentally using linear optical means \cite{21}. The experiment is based on the realization \cite{22} that entanglement properties are well captured by the expectation value of a certain Hermitian operator with respect to two copies of a pure state. As such, this method requires measurements on two copies of a state. It also requires CNOT operations. Application of this method to matter qubits (atomic systems) has also been considered \cite{23}.

In this paper we propose a cavity-QED-based scheme of directly
measuring the concurrence of a two-qubit cavity system. The scheme
works for any arbitrary pure state of a two-qubit cavity  system
even when no prior knowledge about the state is given. The scheme derives from
the realization that the concurrence coincides with the two-particle
visibility under suitable interferometric setups, which in turn
derives from previous theoretical investigations
\cite{19,20,24,25,26,27,28,29,30} that revealed complementarity
between one-particle and two-particle interferences and consequently
intimate relations between two-particle interference and the
concurrence.

\setlength{\unitlength}{0.7cm}
\begin{figure}[h]
\begin{picture}(9.5,3)
\newsavebox{\cavitya}
\savebox{\cavitya}(2,2)[bl]{
\multiput(0,0)(0,1.5){2}{
\multiput(0,0)(2,0){2}{\line(0,1){0.5}}}
\multiput(0,0)(0,2){2}{\line(1,0){2}}
\put(0,0){\qbezier(0,0.5)(1,-0.5)(2,0.5)}
\put(0,0){\qbezier(0,1.5)(1,2.5)(2,1.5)}
}
\newsavebox{\dira}
\savebox{\dira}(2,2)[bl]{
\put(1,1){\circle{1.8}}
}
\newsavebox{\atoma}
\savebox{\atoma}(1,1)[bl]{
\put(0.5,0.3){\circle*{0.2}}
\put(0.5,0.5){\circle{1}}
\multiput(0.2,0.3)(0,0.4){2}{\line(1,0){0.6}}
}
\put(0,0.5){\usebox{\atoma}}
\put(1.5,0){\usebox{\cavitya}}
\put(4,0){\usebox{\dira}}
\put(6.5,0){\usebox{\cavitya}}
\put(2.35,2.5){A}
\put(7.35,2.5){B}
\put(2.35,0.8){$\pi$}
\put(4.85,0.8){$\Phi$}
\put(7.15,0.8){$\pi/2$}
\put(1,1){\line(1,0){1}}
\put(3,1){\line(1,0){1.1}}
\put(5.9,1){\line(1,0){1.1}}
\put(8,1){\vector(1,0){1.3}}
\end{picture}
\caption{The scheme for single-particle interference. A two-level atom prepared in its ground state passes successively through cavity A, a dispersive interaction region and cavity B. The interaction times of the atom with cavities A and B are chosen such that they correspond to a $\pi$-pulse and a $\frac{\pi}{2}$-pulse interaction, respectively. The dispersive interaction changes the relative phase of the atomic states by $\Phi$.}
\label{f1}
\end{figure}
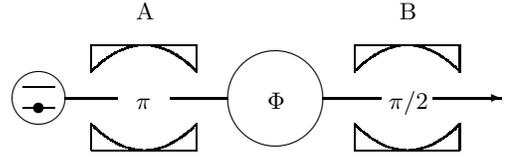

Before describing our proposed scheme, we briefly review the scheme
shown in Fig. \ref{f1} that was considered by Zubairy, et al.
\cite{31} in their investigation of the quantum disentanglement
eraser. In this simple scheme, the concurrence can be directly
measured from the visibility for a specific class of entangled
state.

We consider two cavities A and B that are prepared in an entangled state
\begin{eqnarray}\label{e1}
\alpha|0\rangle_A|1\rangle_B+\beta|1\rangle_A|0\rangle_B,
\end{eqnarray}
where $\alpha=|\alpha| e^{i \theta_a}$ and $\beta
=|\beta|e^{i \theta_b}$ are arbitrary coefficients
with $|\alpha|^2+|\beta|^2=1$. Here $|0\rangle_{A(B)}$ and
$|1\rangle_{A(B)}$ refer to the vacuum and single photon state,
respectively, of cavity A(B).  A two-level atom with the upper level
$|e\rangle$ and the lower level $|g\rangle$ passes successively
through cavity A, a dispersive interaction region and cavity B as
shown in Fig. \ref{f1}. We assume that initially the atom is
prepared in state $|g\rangle$. The interaction time between the atom
and cavity A is chosen such that it corresponds to a $\pi$-pulse
interaction. The interaction then plays the role of the swapping
operation between the atomic state and the state of cavity A. The
dispersive interaction shifts the phase of the atomic state by
$\Phi$ if the atom is in $|e\rangle$. It thus operates as a quantum
phase gate \cite{32,33} for the atom. The interaction time between
the atom and cavity B is chosen such that it corresponds to a
$\frac{\pi}{2}$-pulse interaction. This interaction operates as
$\frac{\pi}{2}$ rotation about X axis for the two states
$|g\rangle|1\rangle_B$ and $|e\rangle|0\rangle_B$. It follows that,
after the atom passes through cavity A, the dispersive interaction
region and cavity B, the final state of the system is given by
\begin{eqnarray}\label{e2}
|\psi\rangle_f &=& \frac{1}{\sqrt{2}}\left(\alpha - \beta e^{i \Phi}\right)|g\rangle|0\rangle_A|1\rangle_B \notag \\
&&-\frac{i}{\sqrt{2}}\left(\alpha + \beta e^{i
\Phi}\right)|e\rangle|0\rangle_A|0\rangle_B.
\end{eqnarray}
From Eq.(\ref{e2}) we find that the probability to find the atom in the upper state $|e\rangle$ is given by
\begin{eqnarray}\label{e3}
P_e=\frac{1}{2}\left(1 + 2 |\alpha \beta| \cos( \Phi-\theta_a +
\theta_b) \right).
\end{eqnarray}
The probability $P_e$ exhibits one-particle interference fringes,
analogous to those of the double-slit pattern, which arise from the
fact that there are two possible paths for the atom to end up in
$|e\rangle$; it can absorb a photon and makes a transition to
$|e\rangle$ in cavity A or in cavity B \cite{31}. The visibility of
this interference pattern is $2|\alpha \beta|$, which coincides with
the concurrence of the initial cavity state of Eq. (\ref{e1}).

The scheme described above is simple and the analysis is easy.
However this scheme is incapable of measuring the concurrence for a
more general entangled state. Below we discuss a somewhat more
complicated setup that is capable of measuring concurrence of an
arbitrary pure entangled state for the two-qubit system.

\setlength{\unitlength}{0.4cm}
\begin{figure}[h]
\begin{picture}(19.5,3)
\newsavebox{\cavityb}
\savebox{\cavityb}(2,2)[bl]{
\multiput(0,0)(0,1.5){2}{
\multiput(0,0)(2,0){2}{\line(0,1){0.5}}}
\multiput(0,0)(0,2){2}{\line(1,0){2}}
\put(0,0){\qbezier(0,0.5)(1,-0.5)(2,0.5)}
\put(0,0){\qbezier(0,1.5)(1,2.5)(2,1.5)}
}
\newsavebox{\dirb}
\savebox{\dirb}(2,2)[bl]{
\put(1,1){\circle{1.8}}
}
\newsavebox{\ramseyb}
\savebox{\ramseyb}(2,2)[bl]{
\multiput(0.2,0)(1.6,0){2}{\line(0,1){1.6}}
\multiput(0,1.6)(1.8,0){2}{\line(1,0){0.2}}
\put(0.2,0){\line(1,0){1.6}}
\put(0,1.6){\line(5,2){1}}
\put(1,2){\line(5,-2){1}}
}
\newsavebox{\atomb}
\savebox{\atomb}(1,1)[bl]{
\put(0.5,0.3){\circle*{0.2}}
\put(0.5,0.5){\circle{1}}
\multiput(0.2,0.3)(0,0.4){2}{\line(1,0){0.6}}
}
\put(1,0){\usebox{\ramseyb}}
\put(3.5,0){\usebox{\dirb}}
\put(6,0){\usebox{\cavityb}}
\put(8.5,0.5){\usebox{\atomb}}
\put(10,0.5){\usebox{\atomb}}
\put(11.5,0){\usebox{\cavityb}}
\put(14,0){\usebox{\dirb}}
\put(16.5,0){\usebox{\ramseyb}}
\put(1.7,2.5){R$_1$}
\put(6.7,2.5){A}
\put(12.2,2.5){B}
\put(17.2,2.5){R$_2$}
\put(1.5,0.8){2$\theta_1$}
\put(4.15,0.8){$\Phi_1$}
\put(6.8,0.8){$\pi$}
\put(12.3,0.8){$\pi$}
\put(14.65,0.8){$\Phi_2$}
\put(17,0.8){2$\theta_2$}
\put(1.2,1){\vector(-1,0){1.2}}
\put(2.8,1){\line(1,0){0.85}}
\put(5.35,1){\line(1,0){1.15}}
\put(7.5,1){\line(1,0){1}}
\put(11,1){\line(1,0){1}}
\put(13,1){\line(1,0){1.15}}
\put(15.85,1){\line(1,0){0.85}}
\put(18.3,1){\vector(1,0){1.2}}
\end{picture}
\caption{The proposed scheme to measure the concurrence of a two-qubit cavity system. Atoms 1 and 2 are prepared in their ground state. They each pass through a cavity, a dispersive interaction region and a Ramsey zone, with the interaction times in cavity and the phase shifts generated in the dispersive interaction region and the interaction time in Ramsey zone as denoted in the figure.}
\label{f2}
\end{figure}
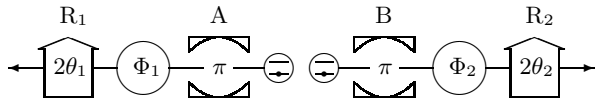

Our proposed scheme is shown in Fig. \ref{f2}. Atom 1(2) in state $|g\rangle_1(|g\rangle_2)$ passes successively through cavity A(B) with a $\pi$-pulse interaction time, a dispersive interaction region changing the relative phase of the atomic states by $\Phi_1(\Phi_2)$, and a Ramsey zone R$_1$(R$_2$) with a $2\theta_1$($2\theta_2$)-pulse interaction time. The initial two-qubit state can be chosen to be any arbitrary state,
\begin{eqnarray}\label{e4}
|\psi\rangle_i = \alpha|0\rangle_{A}|0\rangle_{B}+\beta|0\rangle_{A}|1\rangle_{B}+\gamma|1\rangle_{A}|0\rangle_{B}+\delta|1\rangle_{A}|1\rangle_{B}.
\end{eqnarray}
A scheme to generate an arbitrary two-qubit cavity state was presented in \cite{34}.

Straightforward algebra yields that, after the series of interactions depicted in Fig. \ref{f2}, the final state of the system consisting of atoms 1 and 2 and two cavities A and B becomes
\begin{eqnarray}\label{e5}
|\psi\rangle_f=|0\rangle_A|0\rangle_B && \left( A |g\rangle_1 |g\rangle_2 + B |g\rangle_1 |e\rangle_2 \right. \notag \\
&& +\left. C |e\rangle_1 |g\rangle_2 + D |e\rangle_1 |e\rangle_2  \right)
\end{eqnarray}
where the coefficients $A, B, C$ and $D$ can be expressed in a matrix form as
\begin{eqnarray}\label{e6}
\left(\begin{array}{c}
A\\B\\C\\D\end{array}\right)
=
M_1\otimes M_2 \left(\begin{array}{c}
\alpha\\\beta\\\gamma\\\delta\end{array}\right)
\end{eqnarray}
and the matrix $M_j$($j$=1,2) is a 2$\times$2 matrix which describes the series of interactions that atom $j$ experiences, i.e.,
\begin{eqnarray}\label{e7}
M_j=
\left(\begin{array}{cc}
\cos\theta_j & -i \sin \theta_j\\
-i \sin \theta_j & \cos\theta_j \end{array}\right)
\left(\begin{array}{cc}
1 & 0\\
0 & e^{i \Phi_j} \end{array}\right)
\left(\begin{array}{cc}
1 & 0\\
0 & -i \end{array}\right).
\end{eqnarray}

Following the previous investigations \cite{19,20,25} we consider the corrected joint probability $\overline{P}_{e_1 e_2}$ defined as
\begin{eqnarray}\label{e8}
\overline{P}_{e_1 e_2}=P_{e_1 e_2}-P_{e_1}P_{e_2}+\frac{1}{4}
\end{eqnarray}
where $P_{e_1 e_2}$ is the final joint probability of finding both
atoms in the upper level $|e\rangle_1$ and $|e\rangle_2$ after atom
1 and atom 2 pass through their respective interaction regions, and
$P_{e_1}(P_{e_2})$ is the probability of finding atom 1(2) in the
upper level $|e\rangle_1(|e\rangle_2)$ regardless of the state of
atom 2(1). Since $P_{e_1 e_2}=|D|^2$, $P_{e_1}=|C|^2+|D|^2$ and $P_{e_2}=|B|^2+|D|^2$, we have
\begin{eqnarray}
\overline{P}_{e_1 e_2}&=&|D|^2 -(|C|^2+|D|^2)(|B|^2+|D|^2) + \frac{1}{4} \notag \\
&=& |A|^2|D|^2-|B|^2|C|^2+ \frac{1}{4} \notag \\
&=& (|A||D|-|B||C|)(|A||D|+|B||C|)+\frac{1}{4}.\label{e9}
\end{eqnarray}
For any arbitrary complex coefficients $A$, $B$, $C$ and $D$, we have
\begin{eqnarray}\label{e10}
-|AD-BC|\leq|A||D|-|B||C|\leq|AD-BC|.
\end{eqnarray}
We thus can write
\begin{eqnarray}\label{e11}
-|AD-BC|(|A||D|+|B||C|)+ \frac{1}{4} \leq \overline{P}_{e_1 e_2} \notag \\
\leq |AD-BC|(|A||D|+|B||C|)+ \frac{1}{4}.
\end{eqnarray}
Since $|A||D|\leq\frac{1}{2}(|A|^2+|D|^2)$ and $|B||C|\leq\frac{1}{2}(|B|^2+|C|^2)$, we have
\begin{eqnarray}\label{e12}
|A||D|+|B||C| \leq  \frac{|A|^2+|B|^2+|C|^2+|D|^2}{2} =\frac{1}{2}.
\end{eqnarray}
Substituting inequality (\ref{e12}) into inequality (\ref{e11}), we obtain
\begin{eqnarray}\label{e13}
\frac{1-2|AD-BC|}{4} \leq \overline{P}_{e_1 e_2} \leq \frac{1+2|AD-BC|}{4}.
\end{eqnarray}
Since the system undergoes unitary transformations under our experimental setup of Fig.\ref{f2}, and since the concurrence is preserved under unitary transformations, we have
\begin{eqnarray}\label{e14}
2|AD-BC|=2|\alpha\delta-\beta\gamma|.
\end{eqnarray}
In fact, we obtain, through direct calculation using Eqs.(\ref{e6}) and (\ref{e7}),
\begin{eqnarray}\label{e15}
AD-BC=-e^{i(\Phi_1+\Phi_2)}(\alpha\delta-\beta\gamma).
\end{eqnarray}
Eq.(\ref{e13}) can thus be written as
\begin{eqnarray}\label{e16}
\frac{1-2|\alpha\delta-\beta\gamma|}{4} \leq \overline{P}_{e_1 e_2} \leq \frac{1+2|\alpha\delta-\beta\gamma|}{4}.
\end{eqnarray}
which immediately yields, for the visibility,
\begin{eqnarray}\label{e17}
\mathcal{V}=2\left|\alpha\delta-\beta\gamma \right|.
\end{eqnarray}
Thus the visibility of the two-particle fringes is the same as the concurrence of the initial state of
Eq.(\ref{e4}). Hence, our proposed system of Fig.\ref{f2} provides a
way to measure directly the concurrence of a two-qubit cavity
system.

The actual implementation of our scheme to determine the two-particle visibility requires repeated rounds of experiments, because the maximum and minimum values of the probability $\overline{P}_{e_1 e_2}$ are to be found as the four angles $\theta_1, \theta_2, \Phi_1$ and $\Phi_2$ are varied. One needs to adopt a well-organized search routine to quickly find one set of angles ($\theta_{1 max}, \theta_{2 max}, \Phi_{1 max}, \Phi_{2 max}$) and another set ($\theta_{1 min}, \theta_{2 min}, \Phi_{1 min}, \Phi_{2 min}$) at which the probability $\overline{P}_{e_1 e_2}$ takes on the maximum and minimum values, respectively. Otherwise, the number of runs of experiments that one needs to take may be quite large. If, however, some information about the initial state is given prior to the experiment, then the experiment can be made much less demanding. For example, let us consider the case when we know that all the coefficients $\alpha$, $\beta$, $\gamma$ and $\delta$ of Eq.(\ref{e4}) are real, i.e., when $\alpha=a$, $\beta=b$, $\gamma=c$, $\delta=d$ and $a$, $b$, $c$ and $d$ are real. Through straightforward calculations using Eqs.(\ref{e6}), (\ref{e7}) and (\ref{e9}) with $\alpha=a$, $\beta=b$, $\gamma=c$ and $\delta=d$, we obtain
\begin{eqnarray}\label{e18}
\overline{P}_{e_1 e_2}&=&\left[\left\{ \ 2 (a d + b c) \cos2\theta_1 \cos2\theta_2\right.\right. \notag \\
&&\ +2 (a c - b d) \cos\Phi_2 \cos2\theta_1 \sin2\theta_2 \notag \\
&&\ +2 (a b - c d) \cos\Phi_1 \sin2\theta_1 \cos2\theta_2 \notag \\
&&\ +\left[(a^2+d^2)\cos(\Phi_1+\Phi_2)\right. \notag \\
&&\ \ \ -\left.(b^2+c^2)\cos(\Phi_1-\Phi_2)\right]\sin2\theta_1 \sin2\theta_2\left.\right\} \notag \\
&& \times 2 (a d - b c ) + 1 \left.\right]/4.
\end{eqnarray}
This probability is bound, according to Eq.(\ref{e16}), by
\begin{eqnarray}\label{e19}
\frac{1-2|ad-bc|}{4} \leq \overline{P}_{e_1 e_2} \leq \frac{1+2|ad-bc|}{4}.
\end{eqnarray}
It is then immediately clear that the probability $\overline{P}_{e_1 e_2}$ has the maximum value $\frac{1+2|ad-bc|}{4}$ when $2\theta_1=2\theta_2=\frac{\pi}{2}$ and $\Phi_1=-\Phi_2=\pm\frac{\pi}{2}$ and the minimum value $\frac{1-2|ad-bc|}{4}$ when $2\theta_1=2\theta_2=\frac{\pi}{2}$ and $\Phi_1=\Phi_2=\pm\frac{\pi}{2}$. (Here, $ad>bc$ is assumed. If $ad<bc$, then the angles at which the maximum and minimum occur should be interchanged.) In this case of real coefficients, there is therefore no need to run a search routine. One can fix the angles, for example, at $2\theta_1=2\theta_2=\frac{\pi}{2}$ and $\Phi_1=-\Phi_2=\frac{\pi}{2}$ and run the experiments to find $(\overline{P}_{e_1 e_2})_1$. Another round of experiments should be performed at $2\theta_1=2\theta_2=\frac{\pi}{2}$ and $\Phi_1=\Phi_2=\frac{\pi}{2}$ to obtain $(\overline{P}_{e_1 e_2})_2$. The visibility can then be obtained from the two probabilities $(\overline{P}_{e_1 e_2})_1$ and $(\overline{P}_{e_1 e_2})_2$, because we know that the larger and smaller of the two are the maximum and minimum values, respectively, of the probability $\overline{P}_{e_1 e_2}$.

If, in addition to knowing that the coefficients are real, we know
that $b$=$c$=0, i.e., if we know that the state is given in a
Schmidt-decomposed form,
$a|0\rangle_A|0\rangle_B+d|1\rangle_A|1\rangle_B$, then further
simplification of the experiment is possible. In this case, a simple
calculation yields that the maximum of the probability
$\overline{P}_{e_1 e_2}$ occurs at
$2\theta_1=2\theta_2=\frac{\pi}{2}$ and $\Phi_1+\Phi_2=0$ and the
minimum at $2\theta_1=2\theta_2=\frac{\pi}{2}$ and
$\Phi_1+\Phi_2=\pi$. (Here, $ad>0$ is assumed. If $ad<0$, then the
angles at which the maximum and minimum occur should be
interchanged.) Hence, in this case, one round of experiments can be
performed at $2\theta_1=2\theta_2=\frac{\pi}{2}$ and
$\Phi_1=\Phi_2=0$, followed by another round of experiments at
$2\theta_1=2\theta_2=\frac{\pi}{2}$, $\Phi_1=\pi$ and $\Phi_2=0$.
Since $\Phi_2=0$ in both rounds of experiments, the dispersion
interaction region for atom 2 is not needed.

In conclusion, we have shown that the scheme we propose here allows direct measurement of the concurrence of a two-qubit cavity system. It only involves standard cavity-field-atom ($\pi$-pulse time) interactions corresponding to swapping operations, dispersive interactions corresponding to quantum phase shift operations (Z rotations on the Bloch sphere) and Ramsey zones corresponding to single-qubit rotations (X rotations on the Bloch sphere). These operations have been demonstrated experimentally \cite{32,33,35} and therefore our proposed scheme can be realized within the present cavity-QED technologies.

S.M.L., S.W.J. and H.W.L. were supported by a Grant from the Korea Research
Institute for Standards and Science (KRISS). The research of M.S.Z. was
partially supported by a grant from Qatar National Research Fund
(QNRF).

\end{document}